\documentclass[reprint,superscriptaddress,aps,prd,floatfix,nofootinbib,bibnotes]{revtex4-1}
\usepackage[colorlinks=true,linkcolor=black,citecolor=blue, urlcolor=blue,bookmarks=false]{hyperref}
\hypersetup{breaklinks=true}
\usepackage[utf8]{inputenc}
\usepackage{natbib}
\usepackage{bm,mathtools,cleveref}
\usepackage{array}
\usepackage{epsfig}
\usepackage{amsmath}
\usepackage{amssymb}
\usepackage{multirow}
\usepackage{graphicx,subcaption}
\usepackage{comment}
\usepackage{enumitem}
\graphicspath{{./fig/}}
\usepackage[normalem]{ulem}  
\usepackage[dvips]{color} 

\renewcommand{\sout}{\bgroup \color{red} \ULdepth=-.5ex \ULset}

\newcommand{\vev}[1]{\langle{#1}\rangle}

\newcommand{\qv}{\vec{q}\,}

\begin{document}
\title{$K_{1}^{\pm}$ mesons moving in nuclear matter}

\author{Seokwoo Yeo}
\email{ysw351117@gmail.com}
\affiliation{Department of Physics and Institute of Physics and Applied Physics, Yonsei University, Seoul 03722, Korea}
\author{HyungJoo Kim}
\email{hugokm0322@gmail.com}
\affiliation{International Institute for Sustainability with Knotted Chiral Meta Matter (WPI-SKCM$^2$), Hiroshima University, Higashi-Hiroshima, Hiroshima 739-8526, Japan}
\author{Su Houng Lee}%
\email{suhoung@yonsei.ac.kr}
\affiliation{Department of Physics and Institute of Physics and Applied Physics, Yonsei University, Seoul 03722, Korea}

\date{\today}
\begin{abstract}
Observing the mass shifts of mesons immersed in nuclear matter is interesting, as the changes are expected to shed light on the effects of chiral symmetry breaking on the origin of hadron masses.
At the same time, it is important to understand the momentum dependence of the masses for spin-1 mesons, as the changes manifest differently across the two polarization modes.
Here, the 
mass shifts of $K_{1}^{\pm}$ mesons with finite three-momentum in nuclear medium are studied in the QCD sum rule approach. We find that the mass of $K_{1}^{+}$($K_{1}^{-}$) meson is increased(decreased) by the non-trivial momentum effect in both the transverse and longitudinal modes. Specifically, compared to its rest mass in the nuclear medium, in the transverse mode, the mass of $K_{1}^{+}(K_{1}^{-})$ is observed to shift by +2(-55) MeV, while in the longitudinal mode, the mass shift is +13(-11) MeV, all at a momentum of 0.5 GeV. Exploring the medium modifications of $K_{1}$  meson through kaon beams at J-PARC will provide insights on the partial restoration of chiral symmetry in nuclear matter.
\end{abstract}

\maketitle
\section{Introduction}

Among the various particles that make up the universe, hadrons stand out due to their significant contribution to the mass of observable matter.
However, the masses of the individual quarks that compose a hadron are much smaller than the overall mass of the hadron.   Consequently, understanding the mechanisms behind the generation of hadron mass is crucial
\cite{Hatsuda1985,Brown1991,Hatsuda1992,Leupold2010}.
Non-perturbative effects \cite{hooft1976, Collins1977} and chiral symmetry breaking are key mechanisms contributing to the generation of hadron mass. In particular, the mass difference between chiral partners is attributed to the spontaneous breaking of chiral symmetry \cite{Weinberg1967, Shifman1979}.
Observing the mass change of a hadron immersed in matter is expected to provide a crucial link in understanding the connection between chiral symmetry breaking and the hadron's mass. This is because chiral symmetry is anticipated to be partially and/or fully restored in matter.

Vector mesons were initially of interest, as their dilepton decays were expected to offer insights into modified properties arising from the dense initial stages of proton-nucleus or nucleus-nucleus collisions.
\cite{Pisarski1982,Hatsuda1992}. Subsequently, attention shifted to mesons with small widths, as such properties were found to be crucial prerequisites for particle identification.   Mesons within this group are $\omega$\cite{CBELSATAPS:2005iwc}
, $\phi$\cite{KEK-PS-E325:2005wbm,Gubler:2022imm}, $f_{1}(1285)$\cite{Dickson2016,Gubler2017}, $ K^{*}(892)$ \cite{Hatsuda:1997ev}, and $ K_{1}(1270)$\cite{Song2019}.  
The KEK-PS E325 experiments found a nontrivial reduction in the $\omega$ meson and $\phi$ meson mass in nuclear matter \cite{Naruki2006,Muto2007}. 
Experiments with higher statistics are begin performed at the J-PARC E16 experiment \cite{JPARC:2023quf, Aoki:2023qgl} through  $e^+e^-$ decay and will be complemented by the J-PARC E88 experiment, which aims to measure the $\phi$ meson through its $K^+K^-$ decay.
At the same time, to isolate the effect of chiral symmetry breaking and observe the effect realistically, it is crucial to investigate the mass shift between chiral partners with small widths such as the $K^*$ and $K_1$ with the same charge\cite{Lee:2019tvt}. This is so because they both have small widths and are chiral partners so their mass difference solely relies on changes in the chiral order parameter, irrespective of the medium in which they are immersed \cite{Lee:2023ofg}.

In general, when vector mesons are immersed in a medium at rest, the transverse and longitudinal polarization modes will behave differently when moving with finite momentum.
While these momentum dependencies hinder the observation of mass shift, they are dominated by chirally symmetric effects so that the momentum dependencies will be small for the mass difference between chiral partners\cite{Lee:2023ofg}. 
The momentum dependencies of the different polarization modes were studied for the light vector mesons $(\rho, \omega, \phi)$ in QCD sum rule method\cite{Lee1998,*Lee1998Erratum,Leupold1998PRC,Lee2000}.  Recent revisions for $\phi$ meson were made in \cite{Gubler2014,Kim2020}, and heavy quarkonium was studied in \cite{Kim2023}. 
The different polarization modes of a hadron at a given momentum can, in principle, be observed from the complete angular dependence of decay particles \cite{Park2023}. While such measurements are scheduled in the J-PARC E16 experiments for the $\phi$ meson \cite{Aoki2023}, it is essential to estimate the momentum dependencies of the different polarization modes for various hadrons to offer guidelines for these measurements as well as for future experiments.

In this work, we will investigate the mass shift of the $K_1(1270)$ that propagates with finite three-momentum in nuclear matter, using QCD sum rules. While reconstructing $K_{1}$ and $K^{*}$ from hadronic final states poses challenges such as smearing or signal loss due to interactions with the medium, utilizing J-PARC's kaon beam presents a viable approach to observing these particles in nuclear target experiments \cite{Lee:2019tvt, Song2019,Paryev2021}, given that the $K_1$ meson can be observed from a $K+p$ reaction \cite{Gavillet1978}.  We note that in the nuclear medium, $K_{1}^{+}$ and $K_{1}^{-}$ become non-degenerate due to nucleon-induced charge conjugation invariance breaking. Therefore, different effects could be observed in the nuclear medium depending on their charge states.  To study the different charge states in nuclear matter, we will following the method in \cite{Song2019}.

The paper is organized as follows: Section \ref{formalism} provides a brief overview of the QCD sum rules for the $K_{1}^{\pm}$ meson and displays the newly computed Wilson coefficients. Section \ref{result} presents the maximum values for the mass shifts of $K_{1}^{\pm}$ meson. Conclusions are given in Section \ref{conclusion}. The Appendix contains some essential calculations and mathematical expressions.

\section{Formalism}
\label{formalism}
We will follow the vacuum sum rule and similar notation as used in the work of Song et al. \cite{Song2019}, and generalize it to include the finite three-momentum dependence. The quantity we study in the following is the two-current correlation function,
\begin{align}
    \Pi_{\mu\nu}(q)=i\int d^4x e^{iq\cdot x}\vev{T\left[j_{\mu}(x) j_{\nu}(0) \right]},\label{correlationftn}
\end{align}
where $j_{\mu}^{K_{1}^{-}}=\bar{u}\gamma_{\mu}\gamma_{5}s$ and $j_{\mu}^{K_{1}^{+}}=\bar{s}\gamma_{\mu}\gamma_{5}u$. The currents above are not conserved, leading to contributions from pseudo-scalar mesons to the correlation function. Thus, in the vacuum, we get
\begin{align}
    \Pi_{\mu\nu}(q)=-g_{\mu\nu}\Pi_1(q^2)+q_{\mu}q_{\nu}\Pi_2(q^2). \label{pol1}
\end{align}
Here, $\Pi_2$ has contributions from the spin-0 mesons. Therefore, we will extract the $\Pi_1$(spin-1) part to perform QCD sum rule.
The cases of vacuum and medium with $\qv=0$ are well summarized in \cite{Song2019}. 

When $\vec{q}\neq 0$ in medium, the conserved part of the correlation function in Eq.~(\ref{pol1}) can further be decomposed into the transverse and longitudinal polarization directions with respect to the momentum\cite{Lee1998}. These modes can be obtained by first extracting the transverse mode of Eq.~(\ref{pol1}) $\Pi_{\mu \nu}^T$, explained in Appendix \ref{Appendix-b},  and then using it as follows
\begin{align}
    \Pi_L(\omega^2,\qv^2)=\frac{q^2}{\qv^2}\Pi^T_{00},\\
    \Pi_T(\omega^2,\qv^2)=-\frac{1}{2}\left(\Pi_{\mu}^{T,\mu}+\Pi_L\right).
\end{align}

To clearly observe the $\qv$ dependence, we change the variable from $(w^2,\qv^2)$ to $(Q^2,\qv^2)$, where $Q^2\equiv -w^{2}+\qv^{2}$\cite{Kim2020}. In this case, the $\qv^2$ that is absorbed into $Q^2$ denotes a trivial dependence that does not break Lorentz symmetry. On the other hand, the remaining $\qv^2$ exhibits a non-trivial dependence that violates Lorentz symmetry and represents medium modifications due to $\qv^2$. For spin-1 particles, this non-trivial effect is generally polarization dependent.

From the analytic property of the energy dispersion relation at fixed $\vec{q}^2$, we can derive the relationship between the real and imaginary parts of the correlator in terms of the four momenta and the remaining non-trivial three-momentum. 
\begin{align}
    \text{Re}\Pi_{i}(Q^2,\qv^2)=\frac{1}{\pi}\int_{-\qv^2}^{\infty}ds \frac{\text{Im}\Pi_i(s,\qv^2)}{s+Q^2},\label{dispersion relation}
\end{align}
where $i=L$ or $T$.

To reduce the effects of excited resonances, continuum, and higher dimensional condensates, the Borel transform is introduced as
\begin{align}
    \mathcal{M}(M^{2},\qv^2)=\lim_{\substack{n,Q^{2} \to 0,\\{Q^{2}/n=M^{2}}}}(Q^{2})^{n+1}\frac{1}{n!}\left (-\frac{d}{dQ^{2}}\right )^{n}\Pi(Q^2,\qv^2),
\end{align}
where the parameter $M^2$ is called the Borel mass.

\subsection{Vacuum sum rule}

In the vacuum, $\Pi_L=\Pi_T=\Pi_1$.  
Using the operator product expansion (OPE), the current correlator can be expressed as follows, considering terms up to dimension 6.
\begin{align}
\Pi_1(q^2)=B_{0}Q^{2}\text{ln}\frac{Q^2}{\mu^2}+B_{2}\text{ln}\frac{Q^2}{\mu^2}-\frac{B_4}{Q^2}-\frac{B_6}{Q^4},
\end{align}
where
\begin{align}
     B_0 & =\frac{1}{4\pi^2}\left(1+\frac{\alpha_s}{\pi}\right),\\
     B_2 & =\frac{3m_{s}^{2}}{8\pi^2},\\
     B_4 &=-\varepsilon m_s\left<\bar{u}u\right>_{0}+\frac{1}{12}\left<\frac{\alpha_s}{\pi}G^2\right>_{0},\\
     B_6 & =\frac{32\pi\alpha_s}{81}\left(\left<\bar{u}u\right>_{0}^{2}+\left<\bar{s}s\right>_{0}^{2}\right)+\varepsilon\frac{32\pi\alpha_s}{9}\left(\left<\bar{u}u\right>_{0}\left<\bar{s}s\right>_{0} \right).
\end{align}
Here, $n$ in $B_n$ denotes the mass dimension of the operator, the renormalization scale is taken to be $\mu=1$ GeV, and $\left<\mathcal{O}\right>_0$ denotes the vacuum condensate. The four-quark operators can be divided into chiral symmetry breaking and symmetric parts\cite{Kim:2020zae}.  The chiral symmetry breaking part will change in the nuclear medium as the chiral order parameter. However, while the value of the chiral symmetric part in vacuum can be extracted by combining the sum rules for chiral partners, the medium dependence is not so well known\cite{JKim2022}.  Therefore we will just employ 
a factorization ansatz for the dimension 6 four-quark operators\cite{Shifman1979,Hatsuda1992}.

To highlight the distinction between $K_1$ and $K^*$, we introduce $\varepsilon$ in the following manner.
\begin{align}
    \varepsilon =\left\{\begin{matrix}
1 & \text{for}\; K_{1} \\
-1 & \text{for}\; K^{*} \\
\end{matrix}\right.
\end{align}
It is important to note that the only difference between $K_1$ and $K^*$ correlators is proportional to chiral symmetry breaking operators\cite{JKim2022}.

For input parameters, we use the values given at a renormalization scale of $\mu=1 $ GeV\cite{Kim2020}. $m_q$=4.5 MeV, $m_s$=124.4 MeV\cite{Aoki2022}, $\alpha_s$=0.472\cite{Tanabashi2018}, and $\left<\bar{u}u\right>_0$=$(-0.246 \text{GeV})^3$\cite{Aoki2020}. $\left<\alpha_s /\pi G^2\right>_{0}=0.012$ $\text{GeV}^4$ and $\left<\bar{s}s\right>_{0}=0.8\left<\bar{u}u\right>_0$.

The phenomenological spectral function is employed to determine the imaginary part of the correlator
\begin{align}
    \frac{1}{\pi}\text{Im}\Pi_1(q^2)=\frac{m_{K_{1}}^{4}}{g_{K_{1}}^2}\delta(q^{2}-m_{K_{1}}^{2}) \nonumber \\
    +(B_{0}q^{2}-B_{2})\theta(q^{2}-s_{0}),
\end{align}
where $s_0$ is the continuum threshold.

After using Eq.(\ref{dispersion relation}), by performing the Borel transformation and taking the ratio with its derivative, one obtains the following relation for the mass\cite{Song2019}.
\begin{align}
    & m^{2}_{K_{1}}=M^{2}\frac{2B_{0}E_{2}-B_{2}E_{1}/M^{2}+B_{6}/M^{6}}{B_{0}E_{1}-B_{2}E_{0}/M^{2}-B_{4}/M^{4}-B_{6}/M^6},
\end{align}
where
\begin{align}
    & E_{0}=1-e^{-s_{0}/M^{2}},\\
    & E_{1}=1-\left( 1+\frac{s_0}{M^2} \right) e^{-s_{0}/M^{2}},\\
    & E_{2}=1-\left( 1+\frac{s_0}{M^2}+\frac{s_{0}^{2}}{2M^{4}} \right) e^{-s_{0}/M^{2}}.
\end{align}

If the Borel mass M is too large, the pole dominance is lost, and if M is too small, the contribution from higher dimensional condensates increases. Thus, it is necessary to consider an appropriate region called Borel window. We used the two conditions:
\begin{align}
    & M_{min}^{2}: \left| \frac{B_{4}+B_{6}/M^{2}}{B_{0}M^{4}-B_{2}M^{2}}\right| < 0.15,\\
    & M_{max}^{2}: \left| \frac{B_{0}M^{2}(1-E_{1})-B_{2}(1-E_{0})}{B_{0}M^{2}-B_{2}}\right| < 0.7.    
\end{align}
$M_{\text{min}}^{2}$ is determined to ensure that the power corrections do not exceed 15\% of the perturbative part, while $M_{\text{max}}^{2}$ is set to keep the continuum contribution below 70\% of the perturbative part.

The value for $s_{0}$ is chosen so that the extremum of the Borel curve is close to the physical mass of the $K_{1}$. These prescriptions lead the Borel window to be $0.97\leq M^{2}\leq 2.18$ $\text{GeV}^2$, the continuum threshold $s_0$=2.412 $\text{GeV}^2$, and the overlap strength $F_{K_{1}}\equiv m^{2}_{K{_1}}/g^{2}_{K{_1}}$=0.049 $\text{GeV}^2$.

\subsection{Nuclear medium}
In our calculation of the non-scalar part of the OPE, we focused only on the twist-2  terms, known to be the dominant elements\cite{friman1999}. The odd terms in the OPE contribute with opposite signs for the two charged states, a consequence of nucleon-induced charge symmetry breaking. 

We chose to study the following combinations of the correlators. 
\begin{align}
    \Pi_{L,T}(Q^2,\qv^2)=\Pi^{e}(Q^2,\qv^2)\pm q_{0}\Pi^{o}(Q^2,\qv^2),
\end{align}
here, the `$+$' sign in front of $\Pi^{o}$ represents negative charge states, while the `$-$' sign indicates positive charge states. The OPE for the even and odd parts are
\begin{align}
 \Pi^{e}(Q^2,\qv^2)=B_{0}Q^{2}\text{ln}\frac{Q^{2}}{\mu^{2}}+B_{2}\text{ln}\frac{Q^{2}}{\mu^{2}}-\frac{B_{4}^{*}}{Q^{2}}-\frac{B_{6}^{*}}{Q^{4}},\\
 \Pi^{o}(Q^2,\qv^2)=\frac{1}{2Q^{2}}\left(A_{1}^{u}-A_{1}^{s}\right)\rho-\frac{2m_{N}^{2}}{3Q^{4}}\left(A_{3}^{u}-A_{3}^{s}\right)\rho\nonumber\\
 +\frac{\vec{q}_{t}^{2}}{Q^{6}}2m_{N}^{2}\left(A_{3}^{u}-A_{3}^{s}\right)\rho
\end{align}
where
\begin{align}
&    B_{4}^{*}=-\varepsilon m_{s}\left<\bar{u}u\right>_{\rho}
+\frac{1}{12}\left<\frac{\alpha_{s}}{\pi}G^{2}\right>_{\rho}+\frac{m_{N}}{2}\left(A_{2}^{u}+A_{2}^{s}\right)\rho
\nonumber \\
& m_{N}\alpha_{s}A_{2}^{g}\rho\left(\frac{3}{4\pi}-\frac{1}{3\pi}\text{ln}\frac{Q^{2}}{\mu^{2}} \right)
\;-\frac{\vec{q}_{t}^{2}}{Q^{2}}m_{N}\left(A_{2}^{u}+A_{2}^{s}\right)\rho
\nonumber \\
&m_{N}\alpha_{s}A_{2}^{g}\rho\left(-\frac{7\qv_{t}^{2}}{6\pi Q^{2}}+\frac{2\qv_{t}^{2}}{3\pi Q^{2}}\text{ln}\frac{Q^2}{\mu^{2}}-\frac{2\qv_{l}^{2}}{3\pi Q^{2}} \right),\\
& B_{6}^{*}=\frac{32\pi\alpha_{s}}{9}\left(\varepsilon\left<\bar{u}u\right>_{\rho}\left<\bar{s}s\right>_{\rho} +\frac{\left<\bar{u}u\right>^{2}_{\rho}+\left<\bar{s}s\right>^{2}_{\rho}}{9} \right)\nonumber \\
& -\frac{5}{6}m_{N}^{3}\left(A_{4}^{u}+A_{4}^{s}\right)\rho\;
+\rho m_{N}^{3}\left(A_{4}^{u}+A_{4}^{s}\right)\left( \frac{9\vec{q}^{2}_{t}}{2Q^{2}}-\frac{4\vec{q}^{4}_{t}}{Q^{4}}\right)\nonumber \\
& +\rho m_{N}^{3}\left(A_{4}^{u}+A_{4}^{s}\right)\frac{\vec{q}^{2}_{l}}{Q^{2}}.
\end{align}
For convenience, we have represented the three-momentum dependent terms for both the transverse and longitudinal modes together, denoted as $\vec{q}_{t}$ and $\vec{q}_{l}$, respectively. To obtain three-momentum dependence in the transverse mode, one should take the $\qv_{l}=0$ while keeping the $\qv_{t}$ dependent terms. Similarly, when extracting the terms in the longitudinal mode, one should set the $\qv_{t}=0$ while maintaining the $\qv_{l}$ dependent terms.

To calculate the expectation value with baryon density $\rho$, we have used the linear density approximation $\left<\mathcal{O}\right>_{\rho}\simeq \left<\mathcal{O}\right>_{0}+\rho \left<\mathcal{O}\right>_{N}$ and $\rho$ will be set to the normal nuclear matter density $\rho_{0}$=0.17$\text{fm}^{-3}$. The nucleon matrix elements $\left<\mathcal{O}\right>_{N}$ are given by \cite{Gubler2019}
\begin{align}
    m_{q}\left<\bar{u}u+\bar{d}d\right>_{N}=\sigma_{\pi N}, m_{s}\left<\bar{s}s\right>_{N}=\sigma_{sN},\\
    \left<\alpha_{s}/\pi G^{2}\right>_{N}=\frac{8}{9}(-m_{N}+\sigma_{\pi N}+\sigma{sN}),
\end{align}
where $\sigma_{\pi N}=39.7~ \text{MeV}$\cite{Aoki2020}, $\sigma_{sN}=52.9 ~\text{MeV}$\cite{Aoki2020}, and $m_N$ is the nucleon mass taken to be the isospin averaged value $m_{N}=(0.93827+0.93957)/2$ GeV. The numerical values of twist-2 terms can be estimated using the parton distribution function. Thus, $A_{n}^{q}$ and $A_{n}^{g}$ are defined as
\begin{align}
    & \left<\mathcal{ST}(\bar{q}\gamma_{\mu_{1}}D_{\mu_{2}}\cdots D_{\mu_{n}}q(\mu^2)\right>_{N} \nonumber\\
    & \equiv (-i)^{n-1}A_{n}^{q}(\mu^2)\frac{\mathcal{ST}(p_{\mu_{1}}\cdots p_{\mu_{n}})}{2m_{N}},\\
    & \left<\mathcal{ST}(G^{a}_{\alpha \mu_{1}}D_{\mu_{2}}\cdots D_{\mu_{n-1}}G^{a\alpha}_{\mu_{n}})\right>_{N} \nonumber\\
    & \equiv (-i)^{n-2}A_{n}^{g}(\mu^2)\frac{\mathcal{ST}(p_{\mu_{1}}\cdots p_{\mu_{n}})}{m_N},
\end{align}
where the nucleon four momentum $p_\mu$ is taken to be at rest. $A_{n}^{q}$ and $A_{n}^{g}$ are listed in Table \ref{tab:moments}.

\begin{table}[t]
\begin{tabular}{cc|ccc|cc|cc}
\hline \hline
$A_{1}^{u}$ & $A_{1}^{s}$ & $A_{2}^{u}$ & $A_{2}^{s}$ & $A_{2}^{g}$ & $A_{3}^{u}$ & $A_{3}^{s}$ & $A_{4}^{u}$ & $A_{4}^{s}$   \\ \hline
3 & 0 & 0.784 & 0.053 & 0.367 & 0.2178 & 0.0016 & 0.0945 & 0.00121     \\
\hline \hline
\end{tabular}
\caption{$A^q$ and $A^g$ values used in the present work. All values are given at a renormalization scale of 1 GeV\cite{Gubler2019}.}
\label{tab:moments}
\end{table}

$\Pi^e$ and $\Pi^o $ are even and odd under charge conjugation symmetry.  Therefore, the charge even and odd states will become non-degenerate in the medium.  In \cite{Song2019}, we have introduced a sum rule that at least isolates the ground state to either $K_1^-$ or $K_1^+$ state in the spectral density.  This is accomplished through the following sum rule. 
Therefore, in nuclear matter with $\vec{q}$, the dispersion relation Eq.(\ref{dispersion relation}) becomes
\begin{align}
    & \text{Re}\left(
\Pi^{e}\mp \sqrt{m_{K_{1}^{\mp}}^{2}+\vec{q}^{2}}\Pi^{o}\right). \nonumber \\
    & =\frac{2}{\pi}\int\frac{\text{Im}\left(\Pi^{e}\mp \sqrt{m_{K_{1}^{\mp}}^{2}+\vec{q}^{2}}\Pi^{o}\right)ds}{s+Q^{2}}.\label{modified dispersion relation}
\end{align}

By introducing corresponding thresholds $s_0^\pm$ and different residues coupling to the $K_1^\pm$ states, as in Ref. \cite{Song2019}, we derive the following imaginary part at finite $\vec{q}$.
\begin{align}
    & \frac{2}{\pi}\text{Im}\left(\Pi^{e}+\sqrt{m^{2}_{K_{1}^{+}}+\vec{q}^{2}}\Pi^{o}\right)\nonumber\\
    &=
\frac{m^{4}_{K_{1}^{-}}}{g^{2}_{K_{1}^{-}}}
\left( 1+\frac{\sqrt{m^{2}_{K_{1}^{+}}+\vec{q}^{2}}}{\sqrt{m^{2}_{K_{1}^{-}}+\vec{q}^{2}}} \right)
\delta\left({q^{2}-m^{2}_{K_{1}^{-}}}\right)\nonumber \\
& +\left(B_{0}q^{2}-B_{2}\right)\left\{
\left( 1+\frac{\sqrt{m^{2}_{K_{1}^{+}}+\vec{q}^{2}}}{q_{0}}\right)\theta\left(q^{2}-s_{0}^{-}\right)\right. \nonumber\\
& \left. +\left(1-\frac{\sqrt{m^{2}_{K_{1}^{+}}+\vec{q}^{2}}}{q_{0}}\right)\theta\left(q^{2}-s_{0}^{+}\right)
\right\},\nonumber \\
& \frac{2}{\pi}\text{Im}\left(\Pi^{e}-\sqrt{m^{2}_{K_{1}^{-}}+\vec{q}^{2}}\Pi^{o}\right)\nonumber \\
&=
\frac{m^{4}_{K_{1}^{+}}}{g^{2}_{K_{1}^{+}}}
\left( 1+\frac{\sqrt{m^{2}_{K_{1}^{-}}+\vec{q}^{2}}}{\sqrt{m^{2}_{K_{1}^{+}}+\vec{q}^{2}}} \right)
\delta\left({q^{2}-m^{2}_{K_{1}^{+}}}\right)\nonumber \\
& +\left(B_{0}q^{2}-B_{2}\right)\left\{
\left( 1+\frac{\sqrt{m^{2}_{K_{1}^{-}}+\vec{q}^{2}}}{q_{0}}\right)\theta\left(q^{2}-s_{0}^{+}\right)\right. \nonumber \\
& \left.
+\left(1-\frac{\sqrt{m^{2}_{K_{1}^{-}}+\vec{q}^{2}}}{q_{0}}\right)\theta\left(q^{2}-s_{0}^{-}\right)
\right\}.
\end{align}

We now allow for the parameters to change to leading order in  $\rho$ as follows.
\begin{align}
    & F_{\pm}=F_{K_{1}}+F^{'}_{\pm}\rho,\nonumber\\
    & m_{\pm}=m_{K_{1}}+m^{'}_{\pm}\rho, \nonumber\\
    & s_{0}^{\pm}=s_{0}+s_{0}^{'\pm}\rho,
\end{align}
where $m_{\pm}$ and $F_{\pm}$ represent $m_{K_{1}^{\pm}}$ and $ m^{2}_{K_{1}^{\pm}}/g^{2}_{K_{1}^{\pm}}$, respectively. The density independent terms are equivalent to those in the vacuum.
Then, by conducting the Borel transformation and focusing on terms linear in $\rho$, Eq. (\ref{modified dispersion relation}) becomes:
\begin{align}
    &F(M^2)F^{'}_{\pm}+M(M^2,\qv^2)m^{'}_{\pm}+S(M^2,\qv^2)s_{0}^{'\pm}\nonumber\\
    &=C_{\pm}(M^2,\qv^2).\label{linear density equation}
\end{align}
The explicit forms of the functions in Eq.(\ref{linear density equation}) are listed in the Appendix \ref{Appendix-a}.

From Eq.(\ref{linear density equation}), we define the following function: 
\begin{align}
    &V_{\pm}(F^{'}_{\pm},m^{'}_{\pm},s_{0}^{'\pm})\nonumber\\
    &\equiv \int_{M^{2}_{min}}^{M^{2}_{max}}\left\{F(M^2)F^{'}_{\pm}+M(M^2,\qv^2)m^{'}_{\pm}\right.\nonumber \\
    &\left.+S(M^2,\qv^2)s_{0}^{'\pm}-C_{\pm}(M^2,\qv^2)\right\}^{2}dM^{2},\label{V funtion}
\end{align}
We chose the Borel window to be the same as in the vacuum. We aim to identify $F^{'}_{\pm}$, $m^{'}_{\pm}$, and $s_{0}^{'\pm}$ that minimize $V_{\pm}$, the conditions of which are as follows:
\begin{align}
    \frac{\partial V_{\pm}}{\partial F^{'}_{\pm}}=\frac{\partial V_{\pm}}{\partial m^{'}_{\pm}}=\frac{\partial V_{\pm}}{\partial s^{'\pm}_{0}}=0.
\end{align}
This approach results in three coupled linear equations.  
By solving these equations, the final results are obtained.

\section{Results and Discussions}
\label{result}
\begin{figure}[t]
    \centering
    \begin{subfigure}[b]{0.9\linewidth}
        \centering
        \includegraphics[width=\textwidth]{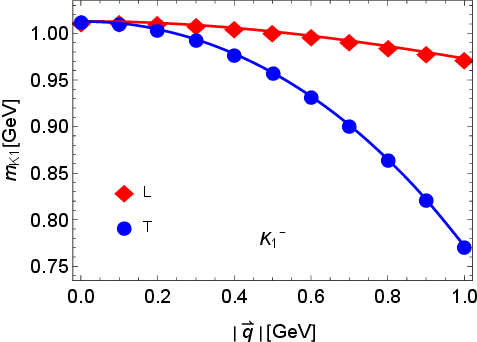}
        \label{fig:sub1}
    \end{subfigure}
    \\
    \begin{subfigure}[b]{0.9\linewidth}
        \centering
        \includegraphics[width=\textwidth]{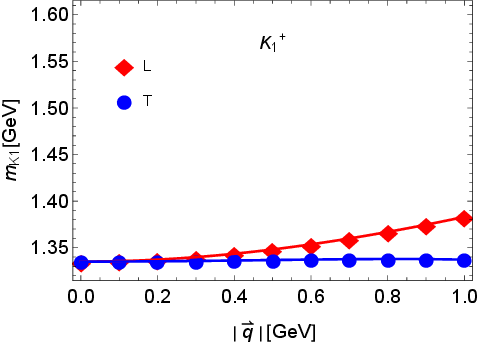}
        \label{fig:sub2}
    \end{subfigure}
    \caption{The masses of $K_{1}^{-}$ and $K_{1}^{+}$ as functions of $\left| \vec{q} \right|$ at normal nuclear matter density. The T (L) means the transverse (longitudinal) mode.}
    \label{fig:test}
\end{figure}

The key finding of our work is the mass shift at finite three-momenta in the nuclear medium. We plot the mass changes of $K_{1}^{+}$ and $K_{1}^{-}$ up to $\left|\qv\right|=1.0$ GeV in Fig.\ref{fig:test}. Our study indicates that $K_{1}^{+}$ shows an increase in mass in both the transverse and longitudinal modes, notably in the longitudinal mode. On the other hand, $K_{1}^{-}$ exhibits a mass reduction across both modes, with a larger effect in the transverse mode. In particular, at a momentum of 0.5 GeV, the mass of $K_{1}^{+}$ is shifted by +2 MeV in the transverse mode and +13 MeV in the longitudinal mode, while $K_{1}^{-}$ demonstrates a shift of -55 MeV in the transverse mode and -11 MeV in the longitudinal mode. For both $K_{1}^{\pm}$ mesons, the direction of mass shifts at finite three momenta aligns with the mass shift directions due to the density at $\vec{q}=0$.
Also, we re-evaluated the mass changes without three-momentum in the nuclear matter reported in \cite{Song2019}, we find the shift for $K_{1}^{-}$ and $K_{1}^{+}$ to be -257 MeV +65 MeV, respectively.

As can be seen in Fig.\ref{fig:test}, in the longitudinal modes, the changes in the two charge states are similar in magnitude but differ in direction.  On the other hand, the transverse mode displays different patterns. From the perspective of the operator product expansion (OPE), $\Pi^{o}$ behaves oppositely across charge states and shows momentum dependence solely in the transverse mode. This leads to the bulk of the observed differences, although on the phenomenological side, the mixing of charge-dependent parameters necessitates a careful evaluation of each term's contribution.

\begin{figure}[t]
    \centering
    \begin{subfigure}[b]{0.9\linewidth}
        \centering
        \includegraphics[width=\textwidth]{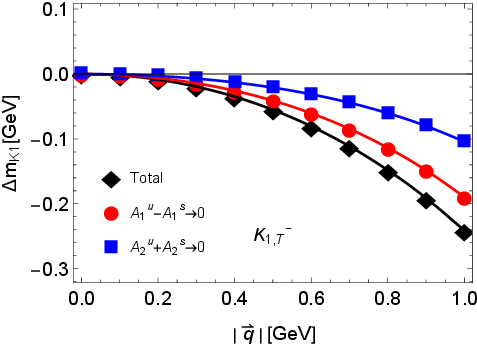}
        \label{fig:sub1}
    \end{subfigure}
    \\
    \begin{subfigure}[b]{0.9\linewidth}
        \centering
        \includegraphics[width=\textwidth]{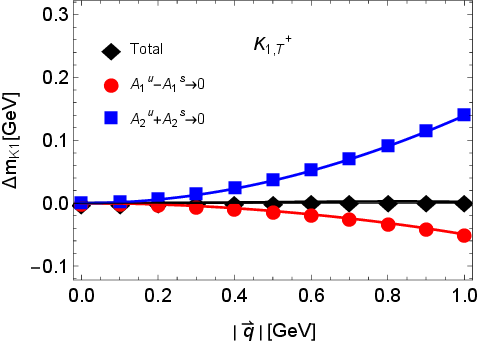}
        \label{fig:sub2}
    \end{subfigure}
    \caption{The effects of the non-scalar operators on $\Delta m_{K_1}(\left| \qv \right|)=m_{K_1}(\left| \qv \right|)-m_{K_1}(0)$ in the transverse mode. The term Total indicates that the full OPE calculated in this study is taken into account. The subscript T in $K_{1,T}^{\pm}$ means the transverse mode.}
    \label{fig:test2}
\end{figure}

\begin{figure}[t]
    \centering
    \begin{subfigure}[b]{0.9\linewidth}
        \centering
        \includegraphics[width=\textwidth]{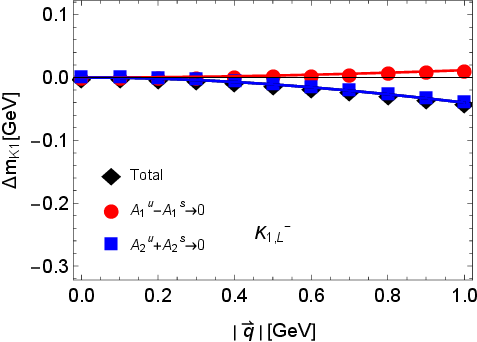}
        \label{fig:sub1}
    \end{subfigure}
    \\
    \begin{subfigure}[b]{0.9\linewidth}
        \centering
        \includegraphics[width=\textwidth]{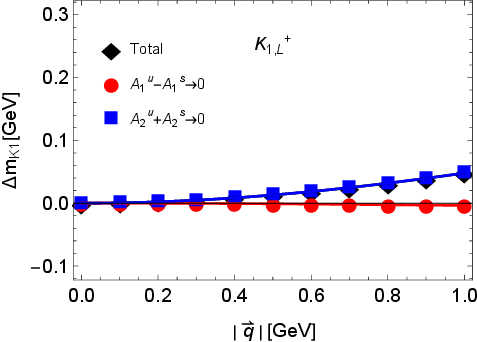}
        \label{fig:sub2}
    \end{subfigure}
    \caption{The effects of the non-scalar operators on $\Delta m_{K_1}(\left| \qv \right|)=m_{K_1}(\left| \qv \right|)-m_{K_1}(0)$ in the longitudinal mode. The term Total indicates that the full OPE calculated in this study is taken into account. The subscript L in $K_{1,L}^{\pm}$ means the longitudinal mode.}
    \label{fig:test3}
\end{figure}

To determine which condensate mostly influences the momentum dependence, we have examined  $\Delta m_{K_1}(\left| \qv \right|)=m_{K_1}(\left| \qv \right|)-m_{K_1}(0)$ after setting each term to zero.
Fig.\ref{fig:test2} illustrates the effects of the two dominant terms in the transverse mode, namely, $A_{1}^{u}-A_{1}^{s}$ and $A_{2}^{u}+A_{2}^{s}$ terms.
Comparing the two different charge states, when $A_{1}^{u}-A_{1}^{s} \rightarrow 0$, the modifications are observed in the same direction, whereas taking $A_{2}^{u}+A_{2}^{s}=0$ leads to changes in the opposite direction.
This indicates that while the $A_{2}^{u}+A_{2}^{s}$ term within $\Pi^{e}$ is indeed responsible for the shift in the same direction, the $A_{1}^{u}-A_{1}^{s}$ term, belonging to $\Pi^{o}$, is responsible for the distinction between $K_{1}^{+}$ and $K_{1}^{-}$ as expected. This is because these operators are proportional to the quark charge with odd numbers of Lorentz indices, similar to vector repulsion in Walecka-type models\cite{Walecka:1974qa}.
In the longitudinal mode, the dominant effect of the $A_{1}^{u}-A_{1}^{s}$ term is seen in Fig.\ref{fig:test3}. Here, the contributions from other terms, including the $A_{2}^{u}+A_{2}^{s}$ term, are small.   Although with lesser effect, $A_{3}^{u}-A_{3}^{s}$ term, which is also part of $\Pi^o$,  has similar effects as the $A_{1}^{u}-A_{1}^{s}$ term, 
leading to shifts in the different directions for the two charge states.

Let us now further illustrate the difference between the longitudinal and transverse modes across the different charge states.  As illustrated in Fig.\ref{fig:test}, it can be seen that both $m_{K_{1}^{\pm},L}-m_{K_{1}^{\pm},T}$, where $L$ and $T$ represent polarization modes, are positive and the difference is large in the negative charge state. This tendency can be understood from the OPE perspective. In the OPE side, the quantities corresponding to $m_{K_{1}^{\pm},L}-m_{K_{1}^{\pm},T}$ are 
\begin{align}
    &\text{Re}(\Pi^{e}\mp \sqrt{m^{2}_{K_{1}^{\mp}}+\qv^2}\Pi^{o})_{L}\nonumber\\
    &-\text{Re}(\Pi^{e}\mp \sqrt{m^{2}_{K_{1}^{\mp}}+\qv^2}\Pi^{o})_{T},
    \label{L-T}
\end{align}
and are given in Eq.(\ref{L-T result}) in Appendix \ref{Appendix-a}. From this we find that the difference between the two charged states is proportional to $A_{3}^{u}-A_{3}^{s}$. In fact, when  $A_{3}^{u}-A_{3}^{s}$ is taken to be zero, $m_{K_{1}^{\pm},L}-m_{K_{1}^{\pm},T}$ exhibit similar values across the two charged states. However, when $A_{2}^{u}+A_{2}^{s}$ is taken to be zero, the difference $m_{K_{1}^{\pm},L}-m_{K_{1}^{\pm},T}$, exhibits  pattern similar to that observed in Fig.\ref{fig:test}. Thus, the OPE perspective offers a clear explanation that the quark charge dependence induces the difference in the charge state, which is in line with the mass difference between different charge states at finite density when $\vec{q}=0$.  

The results of this study provide guidelines for future experiments. The order of the non-trivial momentum-induced mass shift of $K_{1}(1270)$ is similar to that observed in previous studies of other particles, including ($\rho$, $\omega$, $\phi$) \cite{Lee1998,*Lee1998Erratum,Leupold1998PRC,Lee2000,Gubler2014,Kim2020} and heavy quarkonium \cite{Kim2023}. However, in the transverse mode, $K_{1}^{-}$ has a relatively larger momentum dependence. This underlines the necessity of investigating the medium modification of $K_{1}^{-}$ at low momentum or identifying the different polarization modes through the angular dependences of the decay particles\cite{Park:2024vga}.
As detailed in \cite{Lee:2019tvt}, the dominant decay modes for $K_{1}(1270)$ are identified as $K_{1}(1270) \rightarrow K\rho(42\%)$ and $K_{1}(1270) \rightarrow K^{*}\pi(16\%)$. For instance, the production of $K_{1}^{-}$ has been observed in reactions between $K^{-}$ and nucleons \cite{Gavillet1978,Daum1981}. For $K_{1}^{+}$ meson, as discussed in \cite{Paryev2021}, the reaction $K^{-}p\rightarrow K_{1}(1270)^{+}\Xi^{-}$ can be useful. These processes could be explored using the kaon beam at J-PARC \cite{Song2019,Paryev2021}.

While the momentum dependencies differ between the longitudinal and transverse modes, as well as for different charge states, the mass gap between chiral partners only depends on the chiral order parameter\cite{Lee:2023ofg}. So for $K^{*}(892)$, as previously mentioned, it has the same momentum dependent terms as those of $K_{1}(1270)$ from the perspective of OPE. Consequently, the momentum-induced mass shift for $K^{*}(892)$ is expected to be nearly identical to that of $K_{1}(1270)$. 

\section{Conclusions}
\label{conclusion}
We have investigated the mass modifications of $K_{1}^{\pm}$ mesons with finite three-momentum in nuclear matter using QCD sum rule analysis. We find that, due to non-trivial momentum effects, the mass of $K_{1}^{+}$($K_{1}^{-}$) meson is increased(decreased) in both the transverse and longitudinal modes. Specifically, compared to its rest mass in the nuclear medium, $K_{1}^{+}$($K_{1}^{-}$) shows a mass shift of +2(-55) MeV in the transverse mode, while in the longitudinal mode, the shift amounts to +13(-11) MeV, all at a momentum of 0.5 GeV.

Next, we have improved one of the results in \cite{Song2019}, the mass shift at $\qv$=0. For $K_{1}^{+}$($K_{1}^{-}$), the change amounts to +65(-257) MeV. These numerical findings also imply that in nuclear matter, $K_{1}^{-}$\,$(\bar{u}s)$ experiences attraction, whereas $K_{1}^{+}$\,$(\bar{s}u)$ feels repulsion.
This phenomenon can be explained through the quark content of the meson. When it contains a quark (anti-quark), it experiences repulsion (attraction) with the quarks in the medium. This pattern can also be identified in charged kaons\cite{Martin1981,Hyslop1992,Tolos2013,Song2019} and charged D mesons\cite{suzuki2016}.

Meanwhile, the methodology of QCD sum rules focuses only on the bulk property of the spectral function, leading to a scenario where mass shift and width broadening are interrelated\cite{Leupold1998NPA}. As a result, the mass variations determined in this analysis represent the upper limits within the framework of QCD sum rules.

Nevertheless, our study provides direction on how the mass of $K_{1}^{\pm}$ in the medium with momentum would change, which could be validated through several proposed methods\cite{Gavillet1978,Daum1981,Song2019,Paryev2021,Park:2024vga}.

\section*{Acknowledgements}
This work was supported by 
Samsung Science and Technology Foundation under Project 
No. SSTF-BA1901-04, 
and by the Korea National Research Foundation under grant No. 2023R1A2C3003023 and No. 2023K2A9A1A0609492411. The work was also supported by the World Premier International Research Center Initiative (WPI) under MEXT, Japan.

\appendix
\section*{Appendix}
\renewcommand{\thesubsection}{\Alph{subsection}}
\subsection{Some mathematical expressions}
\label{Appendix-a}
Here, we list the explicit forms of the functions presented in Eq.(\ref{linear density equation}) and also show the calculated outcomes in Eq.(\ref{L-T}).
\begin{widetext}
    \begin{align}
        F(M^2)&=-m^{2}_{K_{1}}e^{-m^{2}_{K_{1}}/M^2}, \\
        M(M^2,\qv^2)&=F_{K_{1}}m_{K_{1}}\left(-2+\frac{2m^{2}_{K_{1}}}{M^2}\right)e^{-m^{2}_{K_{1}}/M^2}+\frac{F_{K_{1}}m_{K_{1}}^{3}}{2(m_{K_{1}}^{2}+\qv^{2})}e^{-m_{K_{1}}^{2}/M^2}, \\
        S(M^2,\qv^2)&=\frac{1}{2}\left(1+\frac{\sqrt{m^{2}_{K_{1}}+\qv^{2}}}{\sqrt{s_{0}}+\qv^{2}}\right)\left(B_{0}s_{0}-B_{2}\right)e^{-s_{0}/M^2}, \\
        C_{\pm}(M^2,\qv^2)&=- m_{s}\left<\bar{u}u\right>_{N}+\frac{\alpha_s}{12\pi}\left<G^2\right>_{N}+\frac{m_N}{2}\left(A_{2}^{u}+A_{2}^{s}\right)\pm \sqrt{m_{K_{1}}^{2}+\qv^2}\frac{1}{2}\left(A_{1}^{u}-A_{1}^{s}\right)\nonumber \\
        &+\frac{32\pi\alpha_s}{9M^2}\left( \left<\bar{u}u\right>_{N}\left<\bar{s}s\right>_{0}\pm\varepsilon\left<\bar{u}u\right>_{0}\left<\bar{s}s\right>_{N}+\frac{2}{9}\left(\left<\bar{u}u\right>_{N}\left<\bar{u}u\right>_{0}+\left<\bar{s}s\right>_{N}\left<\bar{s}s\right>_{0}\right)\right)\nonumber \\
        &-\frac{5m_{N}^{3}}{6M^2}\left(A_{4}^{u}+A_{4}^{s}\right)\mp \sqrt{m^{2}_{K_{1}}+\qv^2}\frac{2m_{N}^{2}}{3M^2}\left(A_{3}^{u}-A_{3}^{s}\right)+\frac{\alpha_s}{\pi}A_{2}^{g}m_{N}\left(\frac{3}{4}-\frac{1}{3}\left(\text{ln}M^{2}-\gamma_{E} \right)\right)\nonumber \\
        & +\frac{\alpha_{s}}{\pi}A_{2}^{g}m_{N}\left(-\frac{7\qv_{t}^{2}}{6M^{2}}+\frac{2}{3}\frac{\qv_{t}^{2}}{M^2}\left(\text{ln}M^{2}+1-\gamma_{E} \right)-\frac{2\qv_{l}^{2}}{3M^{2}} \right) \nonumber \\
        &-\frac{\vec{q}_{t}^{2}m_{N}}{M^2}\left(A_{2}^{u}+A_{2}^{s}\right)\pm \sqrt{m^{2}_{K_{1}}+\qv^2}\frac{\vec{q}_{t}^{2}}{M^4}m_{N}^{2}\left(A_{3}^{u}-A_{3}^{s}\right)\nonumber \\
        &+m^{3}_{N}\left(A_{4}^{u}+A_{4}^{s}\right)\left(\frac{9\vec{q}^{2}_{t}}{4M^4}-\frac{2\vec{q}^{4}_{t}}{3M^6}\right)+m^{3}_{N}\left(A_{4}^{u}+A_{4}^{s}\right)\frac{\vec{q}^{2}_{l}}{2M^4}\nonumber \\
        &+m^{'}_{\mp}\left(\frac{F_{K_{1}}m^{3}_{K_{1}}}{2(m^{2}_{K_{1}}+\qv^2)}e^{-m^{2}_{K_{1}}/M^2}\right)\nonumber \\
        &+s^{'\mp}_{0}\left\{\frac{1}{2}\left(-1+\frac{\sqrt{m^{2}_{K_{1}}+\qv^2}}{\sqrt{s_{0}+\qv^2}}\right)\left(B_{0}s_{0}-B_{2}\right)e^{-s_{0}/M^2}\right\}.
    \end{align}
    For the pole of $K_{1}^{\pm}$
    \begin{align}
    \text{Re}(\Pi^{e}\mp &\sqrt{m^{2}_{K_{1}^{\mp}}+\qv^2}\Pi^{o})_{L}-\text{Re}(\Pi^{e}\mp \sqrt{m^{2}_{K_{1}^{\mp}}+\qv^2}\Pi^{o})_{T} \nonumber \\
    =&-\frac{1}{Q^2}\left( \frac{\qv_{t}^{2}}{Q^2}m_{N}(A_{2}^{u}+A_{2}^{s})\rho +m_{N}\alpha_{s}A_{2}^{g}\rho \left(\frac{7\qv_{t}^{2}}{6\pi Q^2}-\frac{2\qv_{t}^{2}}{3\pi Q^2}\text{ln}\frac{Q^2}{\mu}-\frac{2\qv_{l}^{2}}{3\pi Q^2}\right)\right) \nonumber \\
    &-\frac{1}{Q^4}\left( \rho m_{N}^{3}(A_{4}^{u}+A_{4}^{s})\frac{\qv_{l}^{2}}{Q^2}+\rho m_{N}^{3}(A_{4}^{u}+A_{4}^{s})\left( -\frac{9\qv_{t}^{2}}{2Q^2}+\frac{4\qv_{t}^{2}}{Q^4}\right)\right) \nonumber \\
    &\mp\sqrt{m^{2}_{K_{1}^{\mp}}+\qv^2}\frac{\qv_{t}^{2}}{Q^6}2m_{N}^{2}(A_{3}^{u}-A_{3}^{s})\rho.
    \label{L-T result}
\end{align}
\end{widetext}
\subsection{Calculation of Willson coefficient of the twist-2 quark condensate of mass dimension 4}
\label{Appendix-b}
In this section, we present the process to estimate the one of the willson coefficient of non-scalar terms in detail.
\begin{widetext}
    \begin{align}
        \Pi_{\mu\nu}^{T}(q)&=i\int d^{4}x e^{iqx}\left<T\left[\bar{u}(x)\eta_{\nu\sigma}\gamma_{\sigma}\gamma_{5}s(x)\bar{s}(0)\eta_{\mu\delta}\gamma_{\delta}\gamma_{5}u(0)\right]\right>,\nonumber \\
        \Pi_{\mu\nu}^{T,\tau=2,d=4}(q)&=i\sum_{\alpha}\frac{1}{4}\left<\bar{s}\Gamma_{\alpha}D_{\beta}s\right>_{\rho}\frac{\partial }{\partial q_{\beta}}\text{Tr}\left[\Gamma^{\alpha}\eta_{\nu\sigma}\gamma_{\sigma}\gamma_{5}\tilde{S}^{u}(-q)\eta_{\mu\delta}\gamma_{\delta}\gamma_{5}\right]+(u\leftrightarrow s)\nonumber \\
        &=\frac{i}{4}\sum_{\alpha}\left<\bar{s}\Gamma_{\alpha}D_{\beta}s\right>_{\rho}\text{Tr}[\Gamma^{\alpha}\eta_{\nu\sigma}\gamma_{\sigma}\gamma_{5}\not\! q\gamma^{\beta}\not\! q\eta_{\mu\delta}\gamma_{\delta}\gamma_{5}]\frac{1}{q^4}+(u\leftrightarrow s)
    \end{align}
    where $\eta_{\mu\nu}$ is introduced to extract the conserved part and suppress the pseudo-scalar contribution\cite{Reinders1982}.
    \begin{align}
        \eta_{\mu\nu}&\equiv \left(\frac{q_{\mu}q_{\nu}}{q^2}-g_{\mu\nu}\right).
    \end{align}
    In this case $\Gamma_{\alpha}=\gamma_{\alpha}$.\\ 
    First, by concentrating solely on the strange quark operator, the following results are obtained.
    \begin{align}
        \left<\bar{s}\gamma_{\alpha}D_{\beta}s\right>_{\rho}&\simeq \left<\bar{s}\gamma_{\alpha}D_{\beta}s\right>_{0}+\rho \left<\mathcal{ST}(\bar{s}\gamma_{\alpha}D_{\beta}s)\right>_{N}, \\
        \rho \left<\mathcal{ST}(\bar{s}\gamma_{\alpha}D_{\beta}s)\right>_{N}&=-i \rho A_{2}^{s} \frac{m_{N}}{2}\left(v_{\alpha}v_{\beta}-\frac{g_{\alpha\beta}}{4}\right), \\
        \left(v_{\alpha}v_{\beta}-\frac{g_{\alpha\beta}}{4}\right)\text{Tr}[\Gamma^{\alpha}\eta_{\nu\sigma}\gamma_{\sigma}\gamma_{5}\not\! q\gamma^{\beta}\not\! q\eta_{\mu\delta}\gamma_{\delta}\gamma_{5}]&=4g_{\mu\nu}(q^{2}-2q_{0}^{2})+v_{\mu}(8q_{\nu}q_{0}-8q^{2}v_{\nu})+q_{\mu}(8v_{\nu}q_{0}-4q_{\nu}),\\
        \Pi_{L}^{s}&=\frac{q^2}{\qv^2}\Pi_{00}^{s}\propto 4q^2,\\
        \Pi_{T}^{s}&=-\frac{1}{2}\left(\Pi_{\mu}^{\mu,s}+\Pi_{L}^{s}\right)\propto 4q^{2}-8\qv^{2}.
    \end{align}
    Here, the superscripts T, $\tau$, and d in $\Pi_{\mu\nu}$ have been omitted for convenience.
    Calculating for the u-quark also yields the same result. So, we have
    \begin{align}
        \Pi_L&=\frac{-m_{N}}{2Q^2}(A_{2}^{u}+A_{2}^{s})\rho,\\
        \Pi_T&=-\frac{1}{Q^2}\left(\frac{m_{N}}{2}(A_{2}^{u}+A_{2}^{s})\rho-\frac{\qv^2}{Q^2}m_{N}(A_{2}^{u}+A_{2}^{s})\rho)\right).
    \end{align}
    By examining the odd terms, it can be confirmed that the outcomes for the u-quark and s-quark are opposite sign. As a result, when we consider same-flavour mesons, the odd terms are expected to vanish.
\end{widetext}

\bibliographystyle{apsrev4-1}
\bibliography{refs.bib}

\end{document}